\documentclass{ppig}
\usepackage{epsfig}
\usepackage{booktabs}
\usepackage{ucs}
\usepackage[utf8x]{inputenc}

\usepackage{tabularx}
\usepackage{soul}
\usepackage{xcolor}
\usepackage{xspace}
\usepackage{graphicx}
\usepackage[caption=false]{subfig}  
\usepackage[export]{adjustbox}
\usepackage{amsmath}
\usepackage{textcomp}
\usepackage{listings}
\usepackage{cleveref}
\usepackage{algorithm}
\usepackage[noend]{algpseudocode}
\usepackage{balance}
\usepackage{enumitem}
\usepackage{multirow}
\usepackage{colortbl}
\usepackage{balance}
\usepackage{threeparttable}
\usepackage{multicol}
\usepackage{lipsum}

\definecolor{codegreen}{rgb}{0,0.6,0}
\definecolor{codegray}{rgb}{0.5,0.5,0.5}
\definecolor{codepurple}{rgb}{0.58,0,0.82}
\definecolor{backcolour}{rgb}{0.95,0.95,0.92}
\definecolor{gray}{gray}{0.9}
\definecolor{APA_stats}{RGB}{100, 100, 120}

\newcommand{\CR}{Code Readability\xspace}
\newcommand{\ie}{\emph{i.e.,}\xspace}
\newcommand{\eg}{\emph{e.g.,}\xspace}

\usepackage{enumitem}
\usepackage{tcolorbox}

\newcounter{observation}

\title{Assessing Consensus of Developers' Views on \CR}

\author{Agnia Sergeyuk, Olga Lvova, Sergey Titov, \\ \textbf{Anastasiia Serova, Farid Bagirov, Timofey Bryksin}  \\
  JetBrains Research \\
\{agnia.sergeyuk, olga.lvova, sergey.titov\}@jetbrains.com \\
\{anastasiia.serova, farid.bagirov, timofey.bryksin\}@jetbrains.com}

\date{}

\begin{document}
\maketitle
\thispagestyle{empty}

\pagenumbering{arabic} 
\pagestyle{plain}

\begin{abstract}
The rapid rise of Large Language Models (LLMs) has changed software development, with tools like Copilot, JetBrains AI Assistant, and others boosting developers' productivity. However, developers now spend more time reviewing code than writing it, highlighting the importance of \CR for code comprehension. Our previous research found that existing \CR models were inaccurate in representing developers' notions and revealed a low consensus among developers, highlighting a need for further investigations in this field.

Building on this, we surveyed 10 Java developers with similar coding experience to evaluate their consensus on \CR assessments and related aspects. We found significant agreement among developers on \CR evaluations and identified specific code aspects strongly correlated with \CR. Overall, our study sheds light on \CR within LLM contexts, offering insights into how these models can align with developers' perceptions of \CR, enhancing software development in the AI era.
\end{abstract}

\section{INTRODUCTION} \label{sec:introduction}

Large Language Models (LLMs) have seen rapid advancement, particularly in software development applications, where they serve as coding assistants and power tools like Copilot\footnote{Copilot \url{https://github.com/features/copilot}}, JetBrains AI Assistant\footnote{JetBrains AI Assistant \url{https://plugins.jetbrains.com/plugin/22282-jetbrains-ai-assistant}}, Codeium\footnote{Codeium \url{https://codeium.com/}}, and others.

The evolution of AI-supported programming tools is reshaping software development practices --- while AI enhances productivity, developers spend more time reviewing code than writing it~\cite{mozannar2023reading}.
Given that code comprehension time is generally related to \CR ---the easier the code is to read, the less time it takes for the developer to comprehend it---optimizing the programmer's workflow involves providing suggestions from an LLM that align with developers' understanding of \CR.

In academia, \CR is defined as a subjective, mostly implicit human judgment of how easy the code is to understand~\cite{posnett2011,buse2008,scalabrino2016improving}. However, aligning LLMs with developers' understanding of \CR necessitates an explication of developers' notion of what is readable code. 

In our previous research~\cite{sergeyuk2024reassessing}, we studied if existing predictive models of \CR~\cite{posnett2011,scalabrino2018comprehensive,dorn2012general,mi2022} may be a proxy of developers' \CR notion. This study, in addition to defining 12 \CR-related aspects obtained via the Repertory Grid Technique~\cite{EDWARDS2009785}, revealed a weak correlation between current \CR models and developer evaluations, pointing to a significant gap in these models' ability to reflect developers' perspectives on \CR. It underscored the need for developing more accurate \CR metrics and models. We also found that developers' evaluations of \CR were not always consistently aligned with one another. We hypothesize that these results are dictated by the subjectivity of \CR and its aspects, along with other confounding variables. Therefore, we present a work that builds on top of the previous study, presenting the results of a survey we executed to delve deeper into developers' agreement level on \CR and its aspects.

We conducted a survey involving 10 Java developers from the same company, all with similar coding experience. Our aim was to assess whether a group of developers with similar backgrounds would reach a consensus on \CR assessments and related aspects, and which of those aspects are correlated the most with \CR. 
Each developer evaluated the same set of 30 Java code snippets using a 5-point Likert scale, rating code across 13 \CR-related dimensions. 

The results of the study indicate a statistically significant intraclass correlation on \CR and several related metrics. This suggests that there is a degree of agreement among developers regarding what constitutes \CR. We also found a significant correlation of 12 \CR-related aspects evaluations with an assessment of \CR itself. It implies that LLMs could be tailored to \CR notion by adjusting metrics that are stable among developers and strongly related to \CR.

Overall, our work represents an approach to a deeper and at the time more explicit understanding of \CR concept among developers, which is instrumental in the present rapidly evolving AI-centered world of software development.
\section{BACKGROUND} \label{sec:background}

The development of code-fluent LLMs as coding assistants has fundamentally transformed the coding experience. Several research studies were conducted to examine how humans and AI interact in-depth and to understand how LLMs influence programmer behavior during coding activities, \eg~\cite{mozannar2023reading, liang2023understanding, vaithilingam2022expectation,Barke2023Grounded}.

A comprehensive study conducted by researchers from Cambridge and Microsoft informed the understanding of how developers interact with AI tools and how to improve this experience~\cite{mozannar2023reading}. Mozannar and colleagues studied the impact of GitHub Copilot on programmers' behavior during coding sessions. As a result, they identified 12 common programmer activities related to AI code completion systems. The researchers found that developers spend more time reviewing code than writing it. Indeed, approximately 50\% of a programmer's coding time involved interactions with the model, with 35\% dedicated to double-checking suggestions.

The investigation by Carnegie Mellon University researchers underscored the challenges encountered by developers while working with AI coding assistants~\cite{liang2023understanding}. Their survey results suggest that developers mainly use these tools to save time, reduce keystrokes, and recall syntax. However, according to the same survey's results, the generated code is limited in meeting both functional and non-functional requirements. Additionally, developers struggle to comprehend the outputs of LLM due to the code being too long to read quickly.

Previous studies highlighted the importance of aligning \CR models with human notions, reducing coders' time and mental effort to comprehend AI coding assistants' suggestions. Developers need to quickly comprehend the code proposed by an AI coding assistant before integrating it into a project and implementing any changes. A critical aspect of this process is what is commonly referred to as Code Readability --- the ease with which developers can read and understand code. In this notion, \CR forms a perceived barrier to comprehension that developers must overcome to efficiently work with code~\cite{posnett2011,buse2008,scalabrino2018comprehensive}. 

Addressing developers' expectations regarding the readability of model-suggested code may involve various fine-tuning methods. Specifically, in addition to the fine-tuning process itself, when the developer modifies the model's weights and parameters, contrastive~\cite{le2020contrastive} and reinforcement~\cite{lambert2022illustrating} learning are valuable tools for this purpose. Implementing these methods encompasses the definition of a learning objective --- information about what output is ``desirable'' and what is not. Frequently, this objective is formed by annotating the models' outputs or by formulating rules that indicate users' satisfaction with the produced code.

Our previous research~\cite{sergeyuk2024reassessing} explored the potential of existing state-of-the-art \CR models~\cite{posnett2011,dorn2012general,scalabrino2018comprehensive,scalabrino2016improving,mi2018improving,mi2022} to be learning objectives to guide the process of fine-tuning. 

\textbf{Posnett et al.'s Model.} Posnett, Hindle, and Devanbu introduced a Simpler Model of \CR based on three features: Halstead volume, token entropy, and line count, surpassing the performance of Buse and Weimer's earlier model \cite{posnett2011}. They employed forward stepwise refinement for feature selection, manually incorporating features driven by intuition and familiarity with Halstead's software metrics. 

\textbf{Dorn's Model.} Dorn developed a General Software Readability Model, expanding beyond Java to include multiple programming languages \cite{dorn2012general}. Dorn's approach extended beyond syntactic analysis to include structural patterns, visual perception, alignment, and natural language elements, transformed into numerical vectors. This model also outperformed the retrained Buse and Weimer's model, emphasizing the value of using a wider range of code characteristics in readability assessments.

\textbf{Scalabrino et al.'s Model.} Scalabrino, Linares-Vasquez, and Oliveto proposed a Comprehensive Model integrating syntactic, visual, structural, and textual elements of code \cite{scalabrino2018comprehensive}. Their binary \CR classifier with 104 features surpassed previous models, emphasizing the benefit of textual alongside structural and syntactic features.

\textbf{Mi et al.'s Model.} Mi, Hao, Ou, and Ma introduced a deep-learning-based \CR model leveraging visual, semantic, and structural code representations \cite{mi2022}. This model outperformed traditional machine learning models on a combined dataset, showcasing the potential of deep learning in automating \CR evaluation.

In our previous study, we utilized the Repertory Grid technique~\cite{kelly2003psychology} to establish a user-centric understanding of aspects related to \CR. This understanding served as a proxy for a unified perception of \CR among the developers who participated in the consequent survey. During this survey, they assessed code snippets on various readability-related aspects and provided an overall judgment on whether the presented snippet was readable or not. The data from the survey was then used to assess the agreement between the models described above and human evaluations of \CR. Overall, we found 12 readability-related bi-polar code aspects presented in~\Cref{tab:characteristics}.

\begin{table*}[ht]
\centering
\small
\begin{tabular}{@{}p{0.45\linewidth}p{0.45\linewidth}@{}}
\toprule
\textbf{Readable pole} & \textbf{Unreadable pole} \\
\midrule
\rowcolor{gray}
Code is concise & Code is too long \\
Code reads well from top to bottom & While reading, the eyes jump from top \newline to bottom and back up again \\
\rowcolor{gray}
\textasteriskcentered Code is not sufficiently explained and needs \newline additional info to understand what it does & Code is overexplained \\
The goal of the code is clear & The goal of the code is not clear \\
\rowcolor{gray}
Code uses basic, known code patterns & Code looks unfamiliar, non-standard \\
Functionality is separated logically & Code needs refactoring \\
\rowcolor{gray}
Code is flat and linear & Code is overly nested \\
There is one action per line of code & There are multiple actions on one line \\
\rowcolor{gray}
Code uses named constants & Code uses ``magic numbers'' \\
Naming clarifies code functionality & Naming is confusing \\
\rowcolor{gray}
Code conforms to style guides & Code is poorly formatted \\
There is balance in the color blocks & There are huge chunks of color blocks \newline that stand out in a distracting way \\
\bottomrule
\end{tabular}

\begin{tablenotes}
\centering
\footnotesize
    \item \textasteriskcentered This characteristic forms a continuum, being ``Readable'' in the middle and ``Unreadable'' at the extremes.    
\end{tablenotes}
\medskip
\caption{\CR Aspects}
\label{tab:characteristics}
\end{table*}

Our research uncovered discrepancies in the correlation between existing \CR models and its human evaluations. While Scalabrino's model~\cite{scalabrino2018comprehensive} showed a moderate correlation with human assessments, other models like Posnett et al.'s ~\cite{posnett2011}, Dorn's~\cite{dorn2012general}, and Mi et al.'s ~\cite{mi2022} demonstrated weaker correlations. This variation suggests that using these \CR models as learning objectives to fine-tune code-fluent LLMs for improved readability might not be optimal. A more precise model is needed to guide this adjustment process.

Additionally, we found that developers assess \CR inconsistently, highlighting the need for more standardized and validated definitions of \CR.

To address these findings, our current study aims to investigate if confounding variables contributed to the previously observed inconsistency in \CR assessments by developers and if agreement on \CR is achievable. We also seek to identify stable aspects of \CR that could serve as a foundation for defining \CR and forming learning objectives for future LLM adjustments. Specifically, our research questions are as follows:

\textbf{RQ 1.} Do Java developers with similar backgrounds consistently assess \CR and its related aspects?

\textbf{RQ 2.} Do previously elicited code aspects represent \CR?

\section{METHODOLOGY}\label{sec:method}

\subsection{Sample}

The sample was gathered by sending the survey link to the internal channels of JetBrains with the invitation to Java programmers to participate in the study on \CR.

Based on preliminary power analysis, the sample was designed to consist of 10 Java programmers with varying experience levels. Despite their different experience levels, we assume that they share the same understanding of functional and non-functional requirements, as they have actively participated in developing a shared codebase.

All participants in our study are proficient Java developers. We assessed their experience using both subjective and objective measures. Six participants self-assessed as ``Advanced'', indicating \textit{extensive} experience and high proficiency in Java programming. Four participants self-assessed as ``Intermediate'', signifying a \textit{strong} understanding and ability to work on complex projects. The distribution of objective experience measures is presented in Table \ref{tab:experience}.

\begin{table}[h]
\centering
\small
\begin{tabular}{lc}
\toprule
\textbf{Experience} & \textbf{Frequency} \\
\midrule
More than 10 years & 4 \\
9--10 years & 2 \\
5--6 years & 1 \\
3--4 years & 1 \\
1--2 years & 2 \\
\bottomrule
\end{tabular}
\caption{Distribution of Years of Experience}
\label{tab:experience}
\end{table}

\subsection{Materials}

In the current survey, we employed materials gathered in the previous study~\cite{sergeyuk2024reassessing}~---~a set of 30 AI-generated Java code snippets and a rating list of 12 \CR aspects. We justified the reuse of these materials to retest our previous approach and investigate if our data collection and analysis methodology might have influenced the earlier results on the agreement of code readability assessment by humans. Therefore, we maintained consistency by using the same approach and materials but with greater attention to confounding variables and data analysis.

The code snippets represented the readability of outputs generated by LLMs, which is important for us due to the fact that the overarching goal of this research is to enhance the Human-AI Experience. To create snippets, we selected tasks from the Code Golf game\footnote{Code Golf game \url{https://code.golf/}} as prompts for ChatGPT 3.5 Turbo (due to the timing of the study) to generate Java language solutions for these tasks. Subsequently, we ensured that the snippets were meaningful and executable, adhering to a length limit of 50 lines, as defined by previous \CR models examined in prior research.

The primary aim of the list of \CR-related aspects (see~\Cref{tab:characteristics}), which we formulated from in-depth interviews using the Repertory Grid technique,  was to offer consistent guidance to respondents during the evaluation of \CR. This aimed to ensure that developers assessed code uniformly and focused on key aspects related to readability.

\subsection{Data Collection}

On the greeting page of the survey, participants gave their consent and professional background information. After that, they were presented with a random sequence of the same set of 30 Java code snippets. Participants evaluated \CR of the snippet using the list of 12 bipolar characteristics and \CR itself with a five-point Likert scale measuring how much the code leans to the readable or unreadable pole. Participants could take breaks while completing the task, leading to completion times ranging from 40 minutes to 5 hours. Therefore, we believe that the random presentation of tasks and the flexible break schedule mitigated the effects of fatigue on the evaluations.

\subsection{Data Analysis}

To answer \textbf{RQ 1}, we calculated the intraclass correlation coefficient (ICC) to assess the agreement level of developers evaluating \CR and its aspects~\cite{liljequist2019intraclass}.

Additionally, to answer \textbf{RQ 2}, we calculated the Pearson's correlation coefficient~\cite{cohen2009pearson} of \CR-related aspects evaluations with overall \CR scores to see what metrics are related to \CR.

\section{FINDINGS}\label{sec:findings}

\textbf{RQ 1. Do Java developers with similar backgrounds consistently assess \CR and its related aspects?}

The agreement level on assessments of \CR and related code aspects was found to be mostly from moderate to good~\cite{koo2016guideline}. The numerical values of ICC with corresponding Medians are presented in~\Cref{tab:code_aspects}.
The results of our study support the idea that developers of similar backgrounds would agree on evaluations of \CR and its related aspects. 

\begin{table}[ht]
\centering
\small
\begin{tabularx}{\textwidth}{>{\raggedright}p{2.5cm}Xcc}
\toprule
\textbf{Code Aspect} & \textbf{Poles} \textit{(if represented numerically --- from 2 to -2)} & \textbf{ICC} & \textbf{Median} \\
\midrule
Readability & Readable / Unreadable & \textbf{0.78} & 1 \\
\midrule
\rowcolor{gray}
Code Structure & Functionality is separated logically / Code needs refactoring & \textbf{0.81} & 1 \\
Nesting & Code is flat and linear / Code is overly nested & \textbf{0.80} & 2 \\
\rowcolor{gray}
Understandable Goal & The goal of the code is clear / The goal of the code is not clear & \textbf{0.79} & 2 \\
Code Length & Code is concise / Code is too long & \textbf{0.78} & 2 \\
\rowcolor{gray}
Inline Actions & There is one action per line of code / There are multiple actions on one line & \textbf{0.76} & 2 \\
Reading Flow & Code reads well from top to bottom / While reading, the eyes jump from top to bottom and back up again & \textbf{0.75} & 2 \\
\rowcolor{gray}
Sufficient Contextual Info & Code is not sufficiently explained and needs additional info to understand what it does / Code is overexplained & \textbf{0.74} & 0 \\
Code Style & Code conforms to style guides / Code is poorly formatted & \textbf{0.70} & 1 \\
\rowcolor{gray}
Magic Numbers & Code uses named constants / Code uses "magic numbers" & \textbf{0.69} & 0 \\
Naming & Naming clarifies code functionality / Naming is confusing & \textbf{0.67} & 1 \\
\rowcolor{gray}
Code Patterns & Code uses basic, known code patterns / Code looks unfamiliar, non-standard & \textbf{0.53} & 2 \\
Visual Organization & There is balance in the color blocks / There are huge chunks of color blocks that stand out in a distracting way & -0.03 & 0 \\
\bottomrule
\end{tabularx}
\begin{tablenotes}
\centering
\footnotesize
    \item Significant ICC values (p < 0.05) are highlighted in bold. 
\end{tablenotes}
\caption{Agreement on \CR}
\label{tab:code_aspects}
\end{table}

Prior \CR studies show that human annotators exhibited imperfect agreement, with a correlation around .5 with the mean readability score~\cite{buse2008,dorn2012general}. In our previous study~\cite{sergeyuk2024reassessing}, we did not find even this level of agreement. This discrepancy with the current results might be accounted for by the developers' shared backgrounds. In the current study, developers had similar backgrounds, all having experience consistently contributing to a specific codebase. Therefore, their views on some programming conventions are closed. In contrast, the developers in our previous study had a wide range of years of experience and worked at vastly different companies. Additionally, it might be the case that using Kripendorf's alpha with many missing values affected our previous findings, and that effect was mitigated by our data gathering this time. Namely, we avoid missing values in our study by presenting a fixed set of the same snippets to a fixed number of nonrandom raters and calculating ICC on that data.

Findings from the current study support the possibility of aligning LLMs' outputs with users' notions of readability. However, such alignment may be uniquely achievable within a specific company or among a group of developers with close views on various coding practices. 

It is also noteworthy that not all \CR-related aspects have received a significant level of agreement between developers. \textit{Visual Organization} scale, which represents the balance between color blocks in the code snippets, \eg several lines of comments or big arrays, has a nonsignificant level of agreement. Having nonformal feedback from participants, we hypothesize that the wording and concept of this scale were unclear for developers and should be refined in future studies.

\textbf{RQ 2. Do previously elicited code aspects represent \CR?}

\begin{figure}[ht]
    \centering
    \includegraphics[width=.8\textwidth]{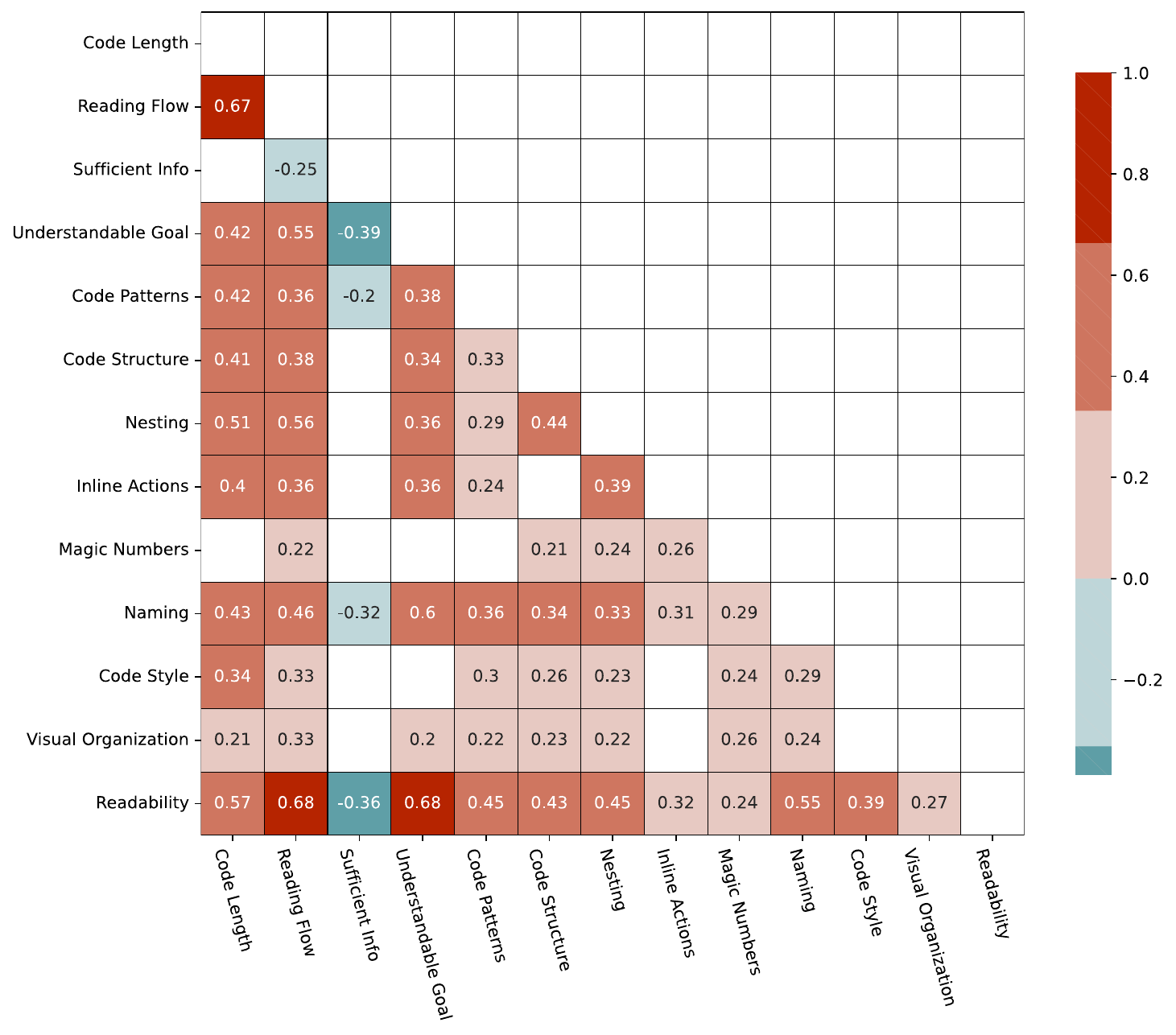}
    \caption{Correlations of \CR-related aspects}
    \label{fig:heatmap}
\end{figure}

The results indicate that aspects of \CR correlate moderately to strongly with the \CR itself. We present a heatmap of statistically significant correlations in~\Cref{fig:heatmap}.

Code aspects from our study, which we identified through in-depth interviews using the Repertory Grids Technique with developers, align with prior research and models of \CR. This alignment, along with the way these aspects were elicited, provides some grounds to hypothesize that they are indeed connected with \CR. Characteristics from our study resemble the combination of structural characteristics with visual, textual, and linguistic features as proposed by later \CR models \cite{dorn2012general, scalabrino2018comprehensive,mi2022}. Moreover, Fakhoury et al. \cite{fakhoury2019improving} investigated commits that were explicitly aimed at \CR-enhancement and observed notable changes in \textit{Complexity, Documentation, and Size} metrics that resemble \textit{Code Structure, Nesting, Sufficient Contextual Info, and Code Length} metrics from our list. In the study of Fakhoury et al., it was also noted that \textit{Code Style and Magic Numbers Usage} are the aspects where improvements in \CR-related commits are prominent. In another study, Peitek et al. \cite{9402005} examined 41 complexity metrics and their influence on program comprehension, discovering that factors such as \textit{Textual Length and Vocabulary Size} increase cognitive load and working memory demand for programmers. 

Further evidence supporting the idea that the code aspects we elicited in our previous study represent developers' notion of \CR  is the statistically significant correlation between the entire list of 12 \CR-related aspects and evaluations of \CR itself. However, there are some differences in the strength of these correlations. The strongest correlation of \CR evaluation is with Naming, Code Length, Understandable Goal, and
Reading Flow metrics. Combined with the fact that Code Length and Understandable Goal are also metrics that gained a good level of agreement among developers who assessed snippets, we can hypothesize that these two code aspects are most representative of \CR and could be used as guidance for LLMs alignment.
\section{CONCLUSION AND FUTURE WORK}\label{sec:conclusion}

This study explored the possibility of agreement among developers on \CR evaluations, with the aim of potentially utilizing \CR as a learning objective for LLMs. Our findings indicate that developers with similar professional backgrounds tend to exhibit a good level of agreement in \CR evaluations. Additionally, certain code aspects related to \CR, \ie \textit{Code Length} and \textit{Understandable Goal}, demonstrate promising potential as representatives of the key scales influencing \CR.

With this supporting evidence in hand, our future endeavors will focus on further exploration of \CR aspects and their potential representations for LLM adjustment with the overarching objective of enhancing user experience with AI assistants in programming.

\bibliography{refs}
\bibliographystyle{apacite} 
\end{document}